\documentclass[runningheads]{llncs}
\usepackage{cite} 
\usepackage{graphicx}
\graphicspath{{./figures/}}
\usepackage{color}
\usepackage{amsmath}
\usepackage[normalem]{ulem}
 
\usepackage{multirow}
\usepackage{booktabs}
\usepackage{subfig}
\usepackage{caption}

\newcommand{\keywords}[1]{\par\addvspace\baselineskip
\noindent\keywordname\enspace\ignorespaces#1}

\begin{document}
\mainmatter  
\title{PGU-net+: Progressive Growing of U-net+ for Automated Cervical Nuclei Segmentation}
\titlerunning{  }
\author{Jie Zhao\inst{1\dagger}, Lei Dai\inst{2\dagger}, Mo Zhang\inst{2}, Fei Yu\inst{2}, Meng Li\inst{2},Hongfeng Li\inst{1}, Wenjia Wang\inst{2}, Li Zhang\inst{1\ddagger}}
\institute{$^{1}$Center for Data Science in Health and Medicine, Peking University, Beijing 100871, China;$^{2}$Center for Data Science, Peking University, Beijing 100871, China; \\
	\  $\dagger$ Joint First Authors, $\ddagger$ Joint Corresponding Authors}
\authorrunning{  }

\maketitle
\begin{abstract}
Automated cervical nucleus segmentation based on deep learning can effectively improve the quantitative analysis of cervical cancer. However, accurate nuclei segmentation is still challenging. The classic U-net has not achieved satisfactory results on this task, because it mixes the information of different scales that affect each other, which limits the segmentation accuracy of the model. To solve this problem, we propose a progressive growing U-net (PGU-net +) model, which uses two paradigms to extract image features at different scales in a more independent way. First, we add residual modules between different scales of U-net, which enforces the model to learn the approximate shape of the annotation in the coarser scale, and to learn the residual between the annotation and the approximate shape in the finer scale. Second, we start to train the model with the coarsest part and then progressively add finer part to the training until the full model is included. When we train a finer part, we will reduce the learning rate of the previous coarser part, which further ensures that the model independently extracts information from different scales. We conduct several comparative experiments on the Herlev dataset. The experimental results show that the PGU-net+ has superior accuracy than the previous state-of-the-art methods on cervical nuclei segmentation.
\end{abstract}
\section{Introduction}
Pap smear is an important test for early screening of precancerous lesions and malignant tumors in gynecology. Accurate segmentation of cervical cancer cells, especially the segmentation of the nuclei, is significant to quantitatively analyze the cervical cancer. Traditional cervical segmentation methods based on image representation are widely used, such as Wavelet \cite{bora2017automated}, support vector machines \cite{tareef2014automated}, template fitting \cite{wu1998parametric}, adaptive thresholding \cite{plissiti2010automated}, genetic algorithms \cite{lassouaoui2003genetic} and graph-cuts \cite{zhang2014segmentation}. Such methods are based on low-level hand-crafted features that usually represent the texture features of the image rather than high-level semantic features. Since the cervical cells of different disease stages undergo global (semantic) changes, if these methods are unable to effectively extract the semantic information of the images, their segmentation accuracy will not satisfy the actual clinical requirements.

The method of deep learning pixel-based object segmentation or detection can simultaneously take into account the characteristic information of different cell structures. The structure of a neural network adjusts the sizes of the receptive fields to adapt to different sizes of targets. Continuous feature extraction through multiple iterations can greatly promote the accuracy of segmentation results. Traditional convolutional neural network U-net \cite{ronneberger2015u} realizes multi-scale information extraction through skip connection. The multi-scale information may have much redundancy and repetition. The use of fixed-size receptive fields for different scale targets is limited to multi-scale learning. Many studies have begun to focus on multi-scale information extraction methods for different target sizes and shapes, such as increasing the receptive field, adding dilated convolution, and merging feature information of different convolution layers, thus improving the classification accuracy of each pixel and generalization of detail features. \cite{song2015accurate} proposed multi-scale convolutional networks and segmentation methods for cervical nucleus and cytoplasm based on graph partitioning. Song et. al. uses a multi-scale deep convolutional neural network to extract diverse feature information and segment overlapping cervical cells \cite{song2016accurate}. The dilated convolution model, which combines multi-scale context information while maintains the receptive field of the original network without losing the resolution of the image space. It has good effects in image classification, target detection and semantic segmentation \cite{yu2017dilated,chen2017deeplab}. However, the dilated rate of the dilated convolution is difficult to design. The artificially designed dilated convolution cannot take into account the characteristic information embodied by the targets of different sizes and shapes. At the same time, learning the feature information of different scales is powerless for the neural network.
\begin{figure*}[!htp]
	\centering
	\centerline{\includegraphics[width=0.6\linewidth]{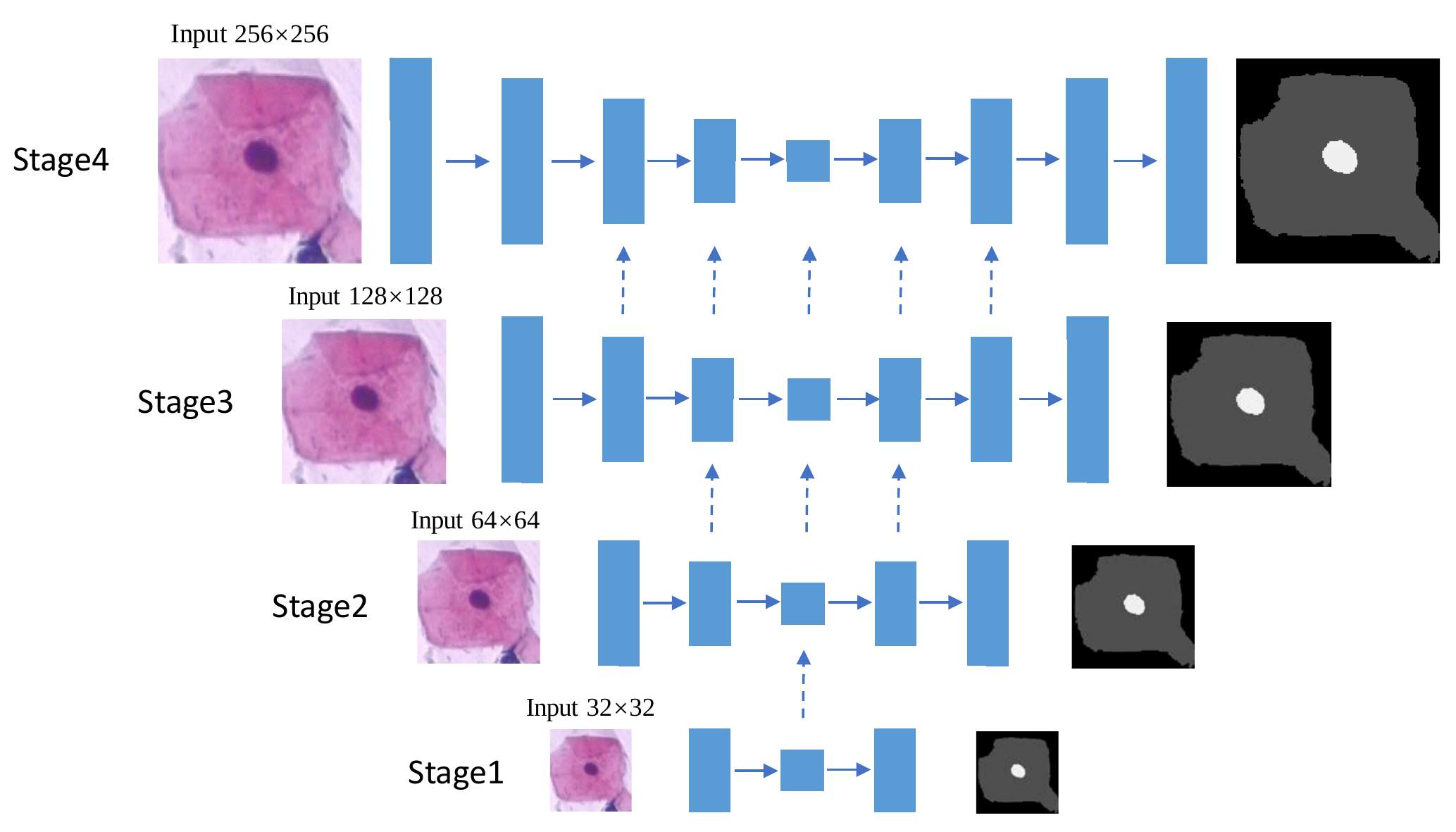}}
	\caption{Flowchart of the experimental procedure. We first use the low-resolution ($32\times32$) image as the input of Stage1, and perform the convolution operation through the solid arrow to get the feature map of each layer (blue boxes). Then we progressively increase the resolution of the input image (the second stage is $64\times64$, the third stage is $128\times128$, and the fourth stage is $256\times256$) and the network is deepened to obtain output results of different sizes. During the stage1 to stage4 process, the middle-layer parameters of the previous stage are continuously transferred (by the dashed arrow).}
	\label{fig:fig_flowchart}
\end{figure*}

To address the aforementioned problems, we propose a novel model - the progressive growing of U-net with residual modules (PGU-net+). Based on the classic U-net, we propose two improvements in the network architecture. First, we added residual modules between different stages (i.e. scales) of the classic U-net. In the first stage with the lowest resolution, we downsample the image and the annotation and train the coarsest part of the model, which learns an approximate shape of the segmentation. We then pass this approximate shape through a residual connection to the next stage with higher resolution, which only learns the residuals of the approximate shape and the annotation (images and annotations will be resampled accordingly in all stages). Thus at each stage, we enforce the model to learn the information related to the current scale. We name this architecture as U-net+. Experiments show that U-net+ can effectively improve the segmentation accuracies.

Second, we adopt a network training paradigm in \cite{karras2017progressive}, called progressive growing. We start to train the model with the coarsest part with downsampled images and annotations, and then progressively add finer part to the training until the full model is included. When training a finer part, we will reduce the learning rate of the previous coarser part, which further encourages the model to extract information from different scales independently. In addition, such paradigm significantly reduces the computational consumption than training the entire model simultaneously. Fig \ref{fig:fig_flowchart} shows the flow chart of this method comprises four stages.
\section{Method}
Classical U-net comprises two major parts: contracting path and expansive path. In the contracting path of deep neural networks, a series of convolution operations can extract feature information to generate coarser feature maps. In the expansive path, corresponding decoding stages progressively recover the resolution of feature maps from coarse to fine.
\subsection{Residual module}
\begin{figure*}[!t]
	\centering
	\centerline{\includegraphics[width=0.25\linewidth]{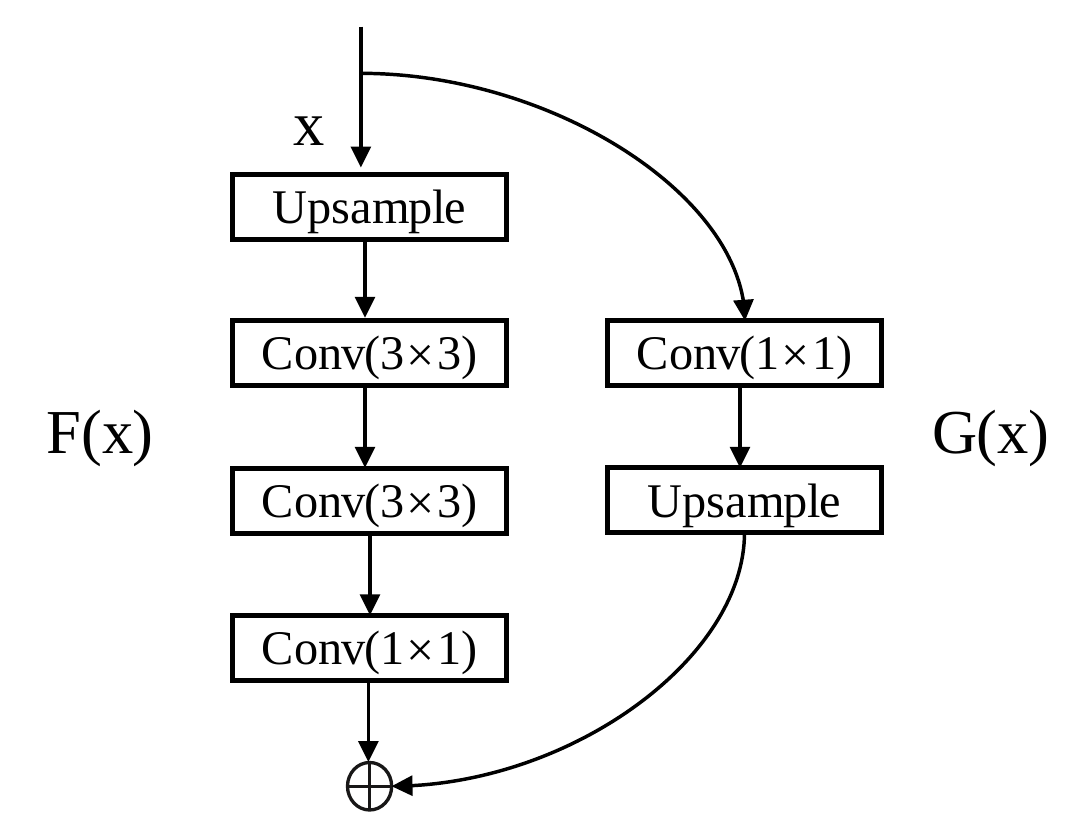}}
	\caption{Residual Module.}
	\label{fig:fig_ResiduleModule}
\end{figure*}

In order to avoid information loss, we introduce a residual module (as shown in Fig \ref{fig:fig_ResiduleModule}) between adjacent scales. The low-resolution feature map of the previous layer is added directly to the high-resolution feature map of the next layer at the pixel level to form residual module. The module is defined as follows:
\begin{equation}
y(p)=F(X(p),W_2(p))+G(X(p),W_2(p))
\label{eq:classic_conv}
\end{equation}
Here $X(p)$ represents the input feature map, $W_1(p)$, $W_2(p)$ denote the weight of the convolution kernel, $y(p)$ represents the output feature map, and the function $F(x,w)$ is the convolution of the expansive path and the maximum pooling operation. $G(x,w)$ represents the residual module. This kind of structure can extract more abundant multi-scale information without increasing the parameters and calculation cost. At each stage, the current network pays more attention to the residual information of adjacent scales to ensure good performance.
\subsection{(Progressive Growing)PG method}
\begin{figure*}[!htp]
	\centering
	\centerline{\includegraphics[width=0.85\linewidth]{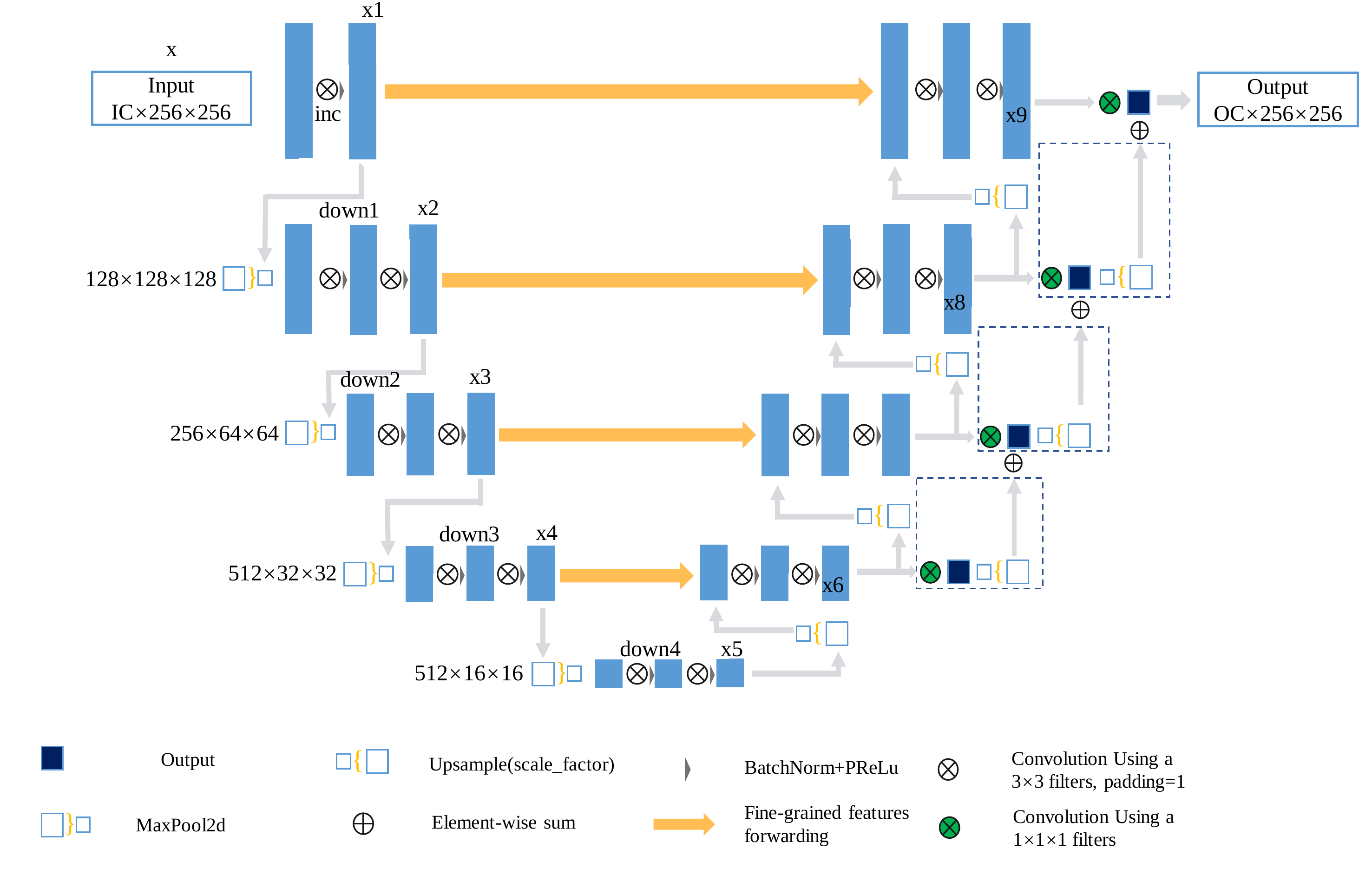}}
	\caption{The U-net structure in the final stage. By migrating the third stage intermediate layer and adding a layer of upsampling and downsampling operations to form the final model structure. In the expansive path, the residual information of the adjacent scale is specially learned, and the input image and the output result size are both $256\times256$.}
	\label{fig:fig_Unet}
\end{figure*}

Traditional convolution kernels or deformable convolution kernels simultaneously learn target information of all scales, which can easily lead to a large number of repetitive or redundant features. If the network is deepened and widened, it will result in high computational and memory cost. Our proposed PGU-net+ model extracts multi-scale feature by introducing a progressive growing \cite{karras2017progressive} training approach. As shown in Fig \ref{fig:fig_flowchart}, we set up 4 training phases. In the first phase, we input a low-resolution image ($32\times32$) to a small U-net network to get the same size of low-resolution output. Then we gradually increase the resolution of the input image to $64\times64$, $128\times128$ and $256\times256$, and continuously add convolution layers to the network to form deeper U-net structures. This type of training allows the network to learn large-scale image coarse structure information first, and then focus on more detailed features at a later stage, rather than learn information of all the scale at the same time. At each stage, the model receives input images of different sizes, so that multi-scale information of target regions of different sizes can be learned step by step. This method makes the model converge faster and have better generalization ability and stability without extra parameters and calculations. Fig \ref{fig:fig_Unet} shows the U-net structure in the final stage with the residual module added to each expansive path. 

We introduce residual module in the extended path of the U-net structure, and adopt a progressive growing training method. At each stage, the model iteratively learns the residual information of adjacent scales. All existing layers in networks remain trainable throughout the training process. When new layers are added to the networks, we adjust smaller learning rate to well-trained, smaller-resolution layers with transferred parameters to avoid sudden shocks on existing networks. By migrating low-resolution image features, the learning of high-resolution images is easier, and the convergence process is faster. The task division of multi-scale learning is further clarified, and the extracted multi-scale information is more accurate and rich.
\section{Experiment and Result}
\subsection{Data Description}
In response to our proposed PGU-net+ structure, this experiment validates our method on the Herlev dataset. The dataset contains 917 images of cervical cancer cells, with each image containing four parts: background, cytoplasm, nucleus and unknown area. Here, we manually determine the unknown area as the background. Considering the difference between large and small nuclei, large and small nuclei are segmented as two types during model training, and all images are normalized to zero mean with unit variance intensity and are resized to a size of $256\times256$.
\subsection{Implementation Details}
We train the model on a single NVIDIA GPU-TITAN. In the first stage, a $32\times32$ raw data is used as input for a small U-net. In the expansive path, the low-resolution feature map is directly doubled and then added to the adjacent high-resolution output to form a residual module, so that the network focuses on learning the residual information of different scales. In the second stage, the original image of $64\times64$ size is used as a U-net input with 2 downsampling and upsampling. In the expansive path, the low resolution feature map is also doubled and added to the adjacent one. And so on into the third and fourth stages. After training 40 epochs at each stage, the next stage is entered. During the parameter transferring process, the learning rate of the trained low-resolution convolutional layer is set to 1e-6, and the newly added convolutional layer learning rate is set to 1e-4 to maintain large-scale feature information and avoid the impact of model changes on existing parameters. We use RMSprop optimization to adaptively adjust the model weights, and the activation function uses RELU.
\begin{figure*}[!t]
	\centering
	\centerline{\includegraphics[width=0.7\linewidth]{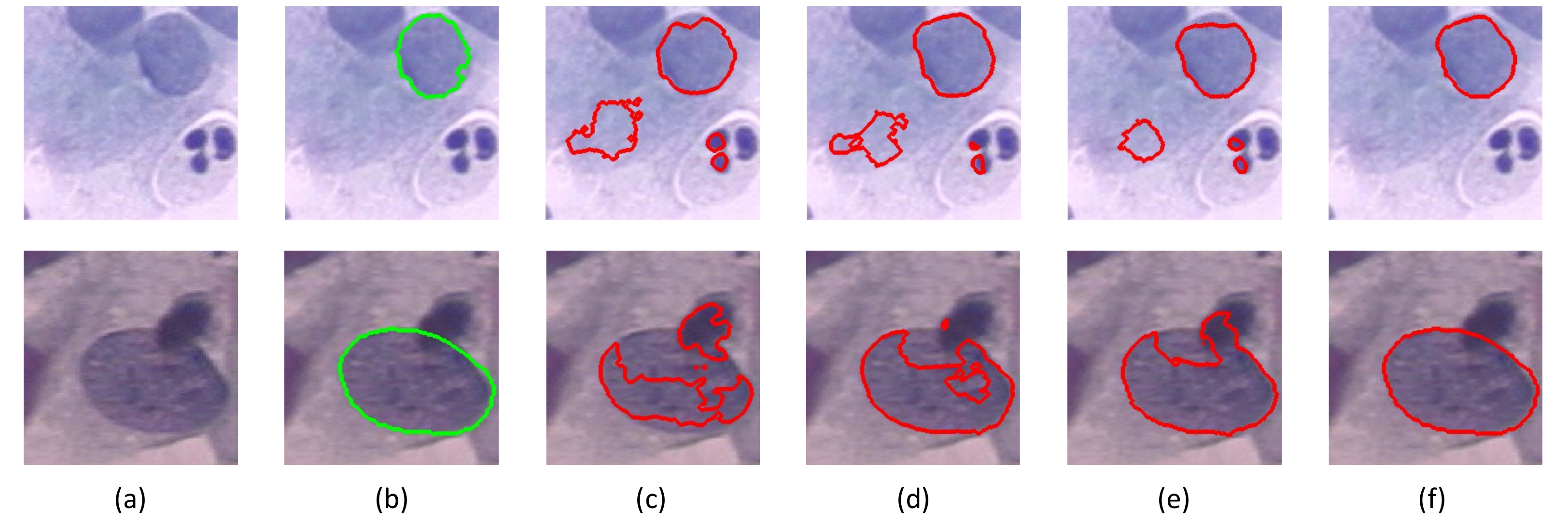}}
	\caption{Examples of the segmentation results. (a) Pap smear images, (b) Manual annotations, (c) Segmentation results of U-net, (d) Segmentation results of U-net+, (e) Segmentation results of PGU-net, (f) Segmentation results of PGU-net+.}
	\label{fig:fig_result}
\end{figure*}
\begin{table}[!tb]
	\begin{center}
		\small
		\resizebox{70mm}{7mm}{
			\begin{tabular*}{0.68\linewidth}{ccccc}
				\hline
				Methods & U-net  & U-Net+ & PGU-net & \textbf{PGU-net+} \\ 
				\hline
				ZSI &0.879$\pm$0.14   & 0.907$\pm$0.10 & 0.911$\pm$0.10 & \textbf{0.926$\pm$0.09} \\
				Precision &0.857$\pm$0.19   & 0.878$\pm$0.14 & 0.890$\pm$0.12 & \textbf{0.901$\pm$0.13} \\
				Recall &0.941$\pm$0.08   & 0.960$\pm$0.07 & 0.950$\pm$0.11 & \textbf{0.968$\pm$0.04} \\
				\hline
		\end{tabular*}}
		\caption{Four sets of experimental results (classical U-net, U-net+, PGU-net, and pgU-net+).)}
		\label{tab:table1}
	\end{center}
\end{table}
\subsection{Experimental Results}
We conduct four sets of experiments. The first group uses a traditional U-net structure to perform nuclear segmentation on $256\times256$ images, including four layers of downsampling and upsampling operations. The second group (short for U-net+) adds a residual module to the expansive path of the traditional U-net structure, making it easier for the training process to grasp features at different scales. The third group (short for PGU-net) applies the progressive growing training method to the traditional U-net structure, continuously increasing the resolution of the input image from 32 to 256 and slowly migrating the low-resolution layer parameters trained in the previous stage. The fourth group adds residual modules in the traditional U-net structure and introduces a progressive growing training mode. The superiority of our proposed PGU-net+ is verified by comparing the four sets of experiments.

By comparing experiments on the Herlev dataset, a total of four set of segmentation results for the dataset are summarized. As shown in Table \ref{tab:table1}, we give three indicators of ZSI, precision and recall. It shows that the U-net network structure with residual module (PGU-net+ model and U-net+ model) is superior to the classic U-net neural network (PGU-net model and U-net model). The progressive growing U-net network structure (PGU-net+ model and PGU-net model) is superior to the classic U-net neural network (U-net+ model and U-net model). The progressive growing with the residual module U-net structure we proposed achieves the best segmentation results. The results of the two groups of cell segmentation experiments are shown in Fig \ref{fig:fig_result}. It can be seen that the PGU-net+ has better segmentation results for cells of different sizes and shapes. We also compare other studies for this dataset. Table \ref{tab:table2} shows the superiority of our model in the three indicators of ZSI, precision and recall under a single model. Our proposed PGU-net+ structure has a segmentation accuracy of 0.925 on the Herlev dataset, and the parameter amount (13M) and computation are much smaller than other models.
\section{Conclusion}
\begin{table}[!tbp]
	\centering
	\small
	\resizebox{70mm}{9.5mm}{
	\begin{tabular*}{0.64\linewidth}{cccc}
		\hline
		Method & ZSI & Precision & Recall \\ 
		\hline
		Unsupervised\cite{gencctav2012unsupervised} & 0.89$\pm$0.15   & 0.88$\pm$0.15 & 0.93$\pm$0.15 \\%
		FCM\cite{chankong2014automatic} &0.80$\pm$0.24   & 0.85$\pm$0.21 & 0.83$\pm$0.25  \\
		SP-CNN\cite{gautam2018cnn} &0.90   & $0.89$ & $0.91$ \\
		DenseUnet\cite{zhao2019automated} &0.91$\pm$0.12   &0.893$\pm$0.14 &0.956$\pm$0.08  \\
		\textbf{Our Method} & \textbf{0.925$\pm$0.09}   & \textbf{0.901$\pm$0.13} & \textbf{0.968$\pm$0.04}  \\%
		\hline
	\end{tabular*}}
	\caption{Comparison of the state-of-the-art methods and proposed method}
	\label{tab:table2}
\end{table}
In this work, we propose to add the residual module in the expansive path of the classic U-net structure, and adopt the progressive growing training mode. Four models (PGU-net+, U-net+,PGU-net and U-net)are used to test on the Herlev dataset. The experimental results show that our model is effective to extract multi-scale information, making the task of extracting multi-scale information more explicit. Furthermore, this residual module can be easily inserted into other higher-order and more complex neural network structures, and the progressive growing training method can also be optimized to solve different scale target detection and target segmentation problems in other fields.
\bibliographystyle{unsrt}
\\\\
{\bf Acknowledgments.} This work is supported in part by the National Key Research and Development Program of China under Grant 2018YFC0910700 and the National Natural Science Foundation of China (NSFC) under Grants 81801778, 11831002, 11701018.
\bibliography{strings,refs}

\begin{thebibliography}{10}

\bibitem{bora2017automated}
Kangkana Bora, Manish Chowdhury, Lipi~B Mahanta, Malay~Kumar Kundu, and
  Anup~Kumar Das.
\newblock Automated classification of pap smear images to detect cervical
  dysplasia.
\newblock {\em Computer methods and programs in biomedicine}, 138:31--47, 2017.

\bibitem{tareef2014automated}
Afaf Tareef, Yang Song, Weidong Cai, David~Dagan Feng, and Mei Chen.
\newblock Automated three-stage nucleus and cytoplasm segmentation of
  overlapping cells.
\newblock In {\em 2014 13th International Conference on Control Automation
  Robotics \& Vision (ICARCV)}, pages 865--870. IEEE, 2014.

\bibitem{wu1998parametric}
Hai-Shan Wu, Joseph Barba, and Joan Gil.
\newblock A parametric fitting algorithm for segmentation of cell images.
\newblock {\em IEEE Transactions on Biomedical Engineering}, 45(3):400--407,
  1998.

\bibitem{plissiti2010automated}
Marina~E Plissiti, Christophoros Nikou, and Antonia Charchanti.
\newblock Automated detection of cell nuclei in pap smear images using
  morphological reconstruction and clustering.
\newblock {\em IEEE Transactions on information technology in biomedicine},
  15(2):233--241, 2010.

\bibitem{lassouaoui2003genetic}
Nadia Lassouaoui and Latifa Hamami.
\newblock Genetic algorithms and multifractal segmentation of cervical cell
  images.
\newblock In {\em Seventh International Symposium on Signal Processing and Its
  Applications, 2003. Proceedings.}, volume~2, pages 1--4. IEEE, 2003.

\bibitem{zhang2014segmentation}
Ling Zhang, Hui Kong, Chien~Ting Chin, Shaoxiong Liu, Zhi Chen, Tianfu Wang,
  and Siping Chen.
\newblock Segmentation of cytoplasm and nuclei of abnormal cells in cervical
  cytology using global and local graph cuts.
\newblock {\em Computerized Medical Imaging and Graphics}, 38(5):369--380,
  2014.

\bibitem{ronneberger2015u}
Olaf Ronneberger, Philipp Fischer, and Thomas Brox.
\newblock U-net: Convolutional networks for biomedical image segmentation.
\newblock In {\em International Conference on Medical image computing and
  computer-assisted intervention}, pages 234--241. Springer, 2015.

\bibitem{song2015accurate}
Youyi Song, Ling Zhang, Siping Chen, Dong Ni, Baiying Lei, and Tianfu Wang.
\newblock Accurate segmentation of cervical cytoplasm and nuclei based on
  multiscale convolutional network and graph partitioning.
\newblock {\em IEEE Transactions on Biomedical Engineering}, 62(10):2421--2433,
  2015.

\bibitem{song2016accurate}
Youyi Song, Ee-Leng Tan, Xudong Jiang, Jie-Zhi Cheng, Dong Ni, Siping Chen,
  Baiying Lei, and Tianfu Wang.
\newblock Accurate cervical cell segmentation from overlapping clumps in pap
  smear images.
\newblock {\em IEEE transactions on medical imaging}, 36(1):288--300, 2016.

\bibitem{yu2017dilated}
Fisher Yu, Vladlen Koltun, and Thomas Funkhouser.
\newblock Dilated residual networks.
\newblock In {\em Proceedings of the IEEE conference on computer vision and
  pattern recognition}, pages 472--480, 2017.

\bibitem{chen2017deeplab}
Liang-Chieh Chen, George Papandreou, Iasonas Kokkinos, Kevin Murphy, and Alan~L
  Yuille.
\newblock Deeplab: Semantic image segmentation with deep convolutional nets,
  atrous convolution, and fully connected crfs.
\newblock {\em IEEE transactions on pattern analysis and machine intelligence},
  40(4):834--848, 2017.

\bibitem{karras2017progressive}
Tero Karras, Timo Aila, Samuli Laine, and Jaakko Lehtinen.
\newblock Progressive growing of gans for improved quality, stability, and
  variation.
\newblock {\em arXiv preprint arXiv:1710.10196}, 2017.

\bibitem{gencctav2012unsupervised}
Asl{\i} Gen{\c{c}}Tav, Selim Aksoy, and Sevgen {\"O}Nder.
\newblock Unsupervised segmentation and classification of cervical cell images.
\newblock {\em Pattern recognition}, 45(12):4151--4168, 2012.

\bibitem{chankong2014automatic}
Thanatip Chankong, Nipon Theera-Umpon, and Sansanee Auephanwiriyakul.
\newblock Automatic cervical cell segmentation and classification in pap
  smears.
\newblock {\em Computer methods and programs in biomedicine}, 113(2):539--556,
  2014.

\bibitem{gautam2018cnn}
Srishti Gautam, Arnav Bhavsar, Anil~K Sao, and KK~Harinarayan.
\newblock Cnn based segmentation of nuclei in pap-smear images with selective
  pre-processing.
\newblock In {\em Medical Imaging 2018: Digital Pathology}, volume 10581, page
  105810X. International Society for Optics and Photonics, 2018.

\bibitem{zhao2019automated}
Jie Zhao, Quanzheng Li, Xiang Li, Hongfeng Li, and Li~Zhang.
\newblock Automated segmentation of cervical nuclei in pap smear images using
  deformable multi-path ensemble model.
\newblock In {\em 2019 IEEE 16th International Symposium on Biomedical Imaging
  (ISBI 2019)}, pages 1514--1518. IEEE, 2019.

\end{thebibliography}
\end{document}